\documentclass[aps,prb,preprint,showpacs,superscriptaddress]{revtex4}
\usepackage{epsfig}
\begin{document}
\topmargin-1.0cm

\title {
Polyfluorene as a model system for space-charge-limited conduction}

\author {M. Arif}
\affiliation {Department of Physics and Astronomy, University of Missouri, Columbia, Missouri 65211}\author {M. Yun}\author {S.
Gangopadhyay}\affiliation {Department of Electrical Engineering, University of Missouri, Columbia, Missouri 65211}\author {K. Ghosh}
\author {L. Fadiga} \affiliation {Department of Physics, Missouri State University, Springfield, Missouri 65804}
\author {F. Galbrecht}\author {U. Scherf}\affiliation {Bergische Universit$\ddot{a}$t Wuppertal, Makromolekulare Chemie, Wuppertal, Germany}\author{S. Guha}\email[Corresponding author E-mail:]{guhas@missouri.edu}
\affiliation {Department of Physics and Astronomy, University of Missouri, Columbia, Missouri 65211}

\date{\today}

\begin{abstract}
Ethyl-hexyl substituted polyfluorene (PF) with its high level of molecular disorder can be described very well by one-carrier
space-charge-limited conduction for a discrete set of trap levels with energy $\sim$ 0.5 eV above the valence band edge. Sweeping the bias above
the trap-filling limit in the as-is polymer generates a new set of exponential traps, which is clearly seen in the density of states
calculations. The trapped charges in the new set of traps have very long lifetimes and can be detrapped by photoexcitation. Thermal cycling the
PF film to a crystalline phase prevents creation of additional traps at higher voltages.

\end{abstract}

 \maketitle
\section{Introduction} \label{sec:intro}
Polyfluorenes (PF) have emerged as an especially attractive conjugated polymer (CP) due to their strong blue emission and excellent electronic
properties, and thus great prospects for device applications.\cite{leclerc,scherf} Such applications rely on charge carrier injection and
transport.\cite{ratner} Recently it was shown the parallel electron mobilities in field-effect transistors based on PF copolymers are one of the
highest: 10$^{-3}$-10$^{-2}$ cm$^2$/Vs. \cite{chua} Charge carrier transport in CPs mainly occurs by variable-range hopping,\cite{parris} where
polarons are the actual carriers,\cite{bussac} and is strongly affected by the presence of traps at the metal-organic interface and in the bulk.
Such trap states (shallow or deep) are favorable energy states; below the conduction band edge these states can capture an electron, and above
the valence band edge can capture a hole.

Impurities and structural defects typically result in discrete trap states within the energy gap.\cite{schmechel} Thermally stimulated currents,
\cite{graupner} photoinduced absorption,\cite{townsend} impedance spectroscopy,\cite{campbell} and current-voltage ({\emph{I-V}) characteristics
in the space-charge-limited current (SCLC) regime\cite{lampert} are some of the experimental methods for the detection of trap states. The
latter yields information on the energy and density of traps as well as charge carrier mobilities. A recent work on polyacenes shows that a
percolative fluctuation process gives a better insight into the electronic conditions determining the crossover from Ohmic to SCLC regime.
\cite{carbone}

SCLC-based models have been used to study trap states in organic molecules and polymers over the last $\sim$40 years. Trapping of carriers in
multi trap levels have been observed in thiophene-based films.\cite {musa} Shallow trapping with a single trap level in naphthalene and
anthracene\cite{pope} was clearly observed in the \emph{I-V } characteristics, which is distinguished by four different regions. These include
(a) the Ohmic region supported by thermal carrier generation, (b) Child's Law in the presence of shallow trapping, (c) trap-filled limit, and
(d) Child's law in the absence of trapping. The above relies on the idea that the energy level of any given trap has a precisely defined value.
Undoubtedly, for single-crystal materials of high chemical and structural purity this is a good approximation for current injection. On the
other hand, electrons or hole traps in amorphous insulators and semiconductors do not have a uniquely defined environment because of the large
structural disorder. As a result isolated discrete energy levels for traps in these materials do not provide an adequate model.

Charge injection in blue-emitting polymers is a topic of heavy debate. In this work we show that charge injection in ethyl-hexyl substituted PF
can be described very well by a one-carrier SCLC-voltage characteristics. To our knowledge this is the first time that charge injection in a
blue-emitting polymer with a high level of molecular disorder is successfully modeled by discrete energy levels for traps. These unexpected
findings suggest that charge injection and transport occur through regions of ordering in the polymer. The initial \emph{I-V} run exhibits a
shallow trap behavior for holes, after which subsequent bias sweeps show a distribution of trap energies in the as-is polymer. The density of
traps $N_t$, their energy levels and the mobility of carriers are obtained from the \emph{I-V} characteristics.
\section{Experimental Details}\label{sec:exptaldetails}

\subsection{PF structure}\label{sec:structure}
Almost all PF derivatives utilize solubilizing side chain substituents anchored at the bridging carbon atom.\cite{tanto} Use of side chains
containing chiral centers creates opportunities for optical activity and the emission of circularly polarized light. Distinct differences in the
optical and electronic properties are observed in two members of the PF family: poly[9,9'-(di n, octyl)fluorene] (PF8) and poly[9,9'-(di
2-ethylhexyl)fluorene] (PF2/6). These polymers are characterized by thermotropic mesophases including a nematic-liquid crystalline (n-LC) phase,
typically above 150 $\rm ^o$C.\cite{chen,knaapila}

PF8 with two linear octyl side chains is characterized by the occurrence of different solid state phases.\cite{chen} Their formation strongly
depends on the conformation of side chains\cite{arif_PRL} and processing conditions that further complicate the discussion of solid state
electronic properties in PF8. PF2/6 with its branched alkyl side chains has a high level of molecular disorder but forms one solid state phase.
In this hexagonal phase the individual PF2/6 chains adopt a five-fold helix (5/2)with coherence lengths exceeding 50 nm. \cite{tanto,knaapila}
Since the application of SCLC technique is based on the purity and crystallinity of a material, PF2/6 serves as a unique ``test'' case because
of its structural properties.

\subsection{Methodology}\label{sec:methodology}
The indium tin oxide (ITO) layer was grown in patterned structures
on glass slides using pulsed laser deposition (PLD).
Poly(ethylenedioxythiophene)-poly(styrenesulfonate) (PEDOT-PSS) was
first spin coated onto the ITO-coated slides (thickness $\sim$ 80
nm) on top of which PF2/6 was spin coated from a toluene solution
(10 mg/ml). This was again coated with a 80 nm PEDOT-PSS layer. The
top Al electrode was deposited by thermal evaporation at a base
pressure of 10$^{-6}$ mbar. These diode structures were then
encapsulated. All fabrication steps were carried out in a nitrogen
glovebox. Light-emitting-diode structures were also fabricated by
directly capping the PF2/6 layer either with Al or with Ca (capped
with Al). The thickness of the PF2/6 layer was 100 nm, device area
was 2-4$\times10^{-4}$ cm$^{2}$ and each sample held 25 devices.
ITO/PEDOT-PSS/(PF2/6)/PEDOT-PSS/Al, ITO/PEDOT-PSS/(PF2/6)/Al, and
ITO/PEDOT-PSS/(PF2/6)/Ca-Al structures are referred to as Sample A,
Sample B, and Sample C, respectively. Sample A is truly a hole-only
device.

We prepared another device (Sample D) where the PF2/6 film was annealed to induce the hexagonal crystalline phase using similar temperature
steps used by Tanto \emph{et al.} in bulk PF2/6.\cite{tanto} After spincoating the PF2/6 layer on PEDOT-PSS coated ITO, the sample was placed in
an oven (inside the glove box) at 80 $\rm^o$C for 15-20 hours. Then the temperature was slowly raised to 150 $\rm^o$C at the rate of 10 $\rm
^o$C/hr and left for two hours at the highest temperature. It was then cooled to room temperature at 1 $\rm^o$C/min. The device structure was
similar to Sample A, with a top PEDOT-PSS layer capped with Al.

The dielectric constant of the polymer (both as-is and annealed) was measured from Al/(PF2/6)/Al  structures by capacitance methods. For both
the as-is and the annealed film the dielectric constant was measured as $\sim$ 2.7 at 10kHz with an RMS voltage of 2.1 V. The $I-V$ measurements
were carried out by a Keithley 236 sourcemeter using manual probes. For the low temperature measurements a CTI closed cycle refrigerator was
used.

\section{Current-voltage characteristics}\label{sec:cv charactersitics}

\subsection{As-is PF2/6 film}\label{sec:asis film}

\begin{figure}
\unitlength1cm
\begin{picture}(4.0,6.5)
\put(-3,-1.){ \epsfig{file=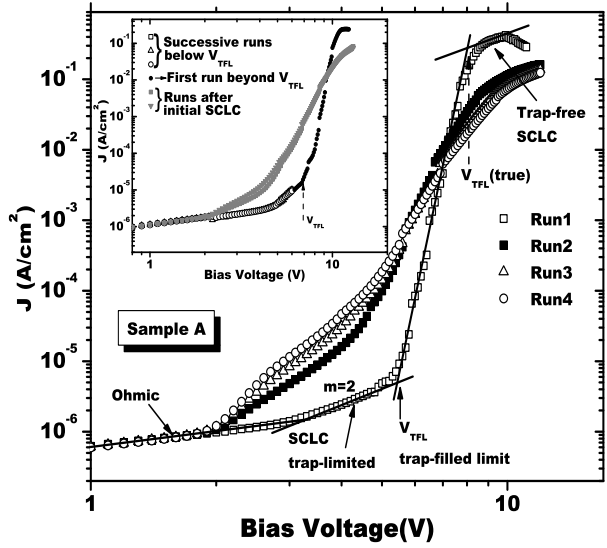, angle=0, width=10.cm, totalheight=8.7cm}}
\end{picture}
\caption{$J-V$ characteristics for sample A. The initial sweep (Run 1) shows all four distinct regions, characteristic of a shallow-trap SCLC
behavior with slope (m)=2 preceding $V_{TFL}$. Subsequent runs show higher values of m. The inset shows $J$ vs. $V$ from another device of the
same sample. The open symbols denote three successive $J-V$ sweeps below $V_{TFL}$; $\bullet$ shows the very first run till trap-free SCLC
voltages. Successive sweeps are shown by the dark gray symbols.} \label{figure1}
\end{figure}

Figure 1 shows the current density ($J$) vs. voltage ($V$) of Sample A for many bias sweeps. The very first measurement clearly shows all four
regions, characteristic of a one carrier single set of traps. These include the (1) Ohmic region, where the current density is proportional to
the voltage, $J=q n_0\mu V/d$, $n_0$ being the free carrier density, \emph{d} the thickness of the polymer layer, and $\mu$ is the carrier
mobility ; (2) SCLC trap-limited region where the current density is given by $J=9 \epsilon \epsilon_0 \mu \Theta V^2/8d^3$, $\Theta$ is the
trapping fraction, $\epsilon$ is the dielectric constant of the polymer, $\epsilon_0$ the permittivity of free space, ; (3) trap-filled limit
where the onset trap-filled voltage $V_{TFL}$ results in the density of traps ($N{_t}$), which is given by
\begin{equation}\label{1}
N_t= \frac{3}{2} \frac{\epsilon \epsilon_0 V_{TFL}}{q d{^2}};
\end{equation}
and (4) the final region is the current only limited by the space-charge and free from the influence of traps. This trap-free SCLC region is
similar to region (2) with $\Theta$ =1. The trapping parameter in the trap-limited conduction region is given by
\begin{equation}\label{2}
\Theta= \frac{N_v}{N_t}\exp(-\frac{E_t}{kT}),
\end{equation}
where we assume the top of the valence band edge is $E_v$ =0; $N_v$ is the density of states within $kT$ of the valence band edge. Since Sample
A is a hole only device the trap-limited current arises from shallow hole traps. This is similar to what is seen in pure organic crystals like
naphthalene.\cite{pope} One can extract all the relevant parameters from the four regions. Region 4 fits very well with $J \propto V^2$; from
the slope we extract the value of $\mu$. (Here we assume field-independent mobilities.) This is then substituted in the trap-limited region (2)
to obtain the value of $\Theta$, and by using Eq. ({\ref{2}}), we evaluate $E_t$. We use $N_v$ = $3\times10^{21}$ cm$^{-3}$.\cite{carbone} Since
the transition region from $V_{TFL}$ to trap-free SCLC is not perfectly vertical (due to reasons discussed later), we use $V_{TFL}$(true), which
is the intersection of trap-free SCLC and the vertical rise, in Eq. ({\ref{1}}).

The $I-V$ curves of Samples B and C are not shown here but are similar to Sample A. Although Ca provides a very low barrier for electrons, the
device is predominantly hole-like. This is most probably because the PLD grown ITO with PEDOT-PSS contact is a better injector of holes than the
metallic ones are for electrons. Single-trap SCLC behavior is observed in the very first run for all sets of devices, and subsequent
measurements show an exponential trap behavior. Table I lists the mobility, trap density, and energy of the traps obtained from the shallow-trap
$J-V$ characteristics for all three samples. The hole mobilities are in the range of 10$^{-5}$ to 10$^{-7}$ cm$^2$/Vs. The trap densities are of
the order of $10^{17} $cm$^{-3}$ and the average trap energy of PF2/6 from all three samples is $\sim$ 0.5 eV.

Our recent work on capacitance-voltage characteristics from PF2/6-based metal-insulator-semiconductor (MIS) diode, where as-is PF2/6 layer was
the active semiconductor layer and Al$_2$O$_3$ was the dielectric layer, yields the concentration of localized charges as 5.7$\times$10$^{17}$
cm$^{-3}$,\cite{yun} similar to values of $N_t$ deduced here. The mobilities extracted in this work from the SCLC model are in good agreement
with other works; time of flight and dark injection space-charge limited current transient measurement from a PF copolymer yields hole
mobilities in the range 10$^{-5}$ to 10$^{-6}$ cm$^2$/Vs.\cite{poplavskyy} Thermally stimulated current (TSC) measurements from doped PF2/6
yield 0.24 eV as the effective hole trap depth; lower trap energies are expected for a doped polymer compared to an undoped one as is the case
in this work. \cite{kadashuchuk}

\begin{table}
\caption{Hole mobility, trap density, and energy of the traps obtained from shallow-trap $J-V$ characteristics.} \label{table1}
\begin{ruledtabular}
\begin{tabular}{cccccc}
Sample   & $\mu$(cm$^2$/Vs) &
$N_t$ (cm$^{-3}$) & $E_t$(eV)\\
\hline A    &  1.8$\times10^{-5}$  & 1.8$\times 10^{17}$  & 0.52$\pm$0.03 \\
B    &  2.5$\times10^{-7}$  & 2.1$\times 10^{17}$ & 0.35$\pm$0.03 \\
C    &  1.9$\times10^{-6}$  & 1.5$\times 10^{17}$ & 0.47 $\pm$0.04\\
\end{tabular}
\end{ruledtabular}
\end{table}

For all devices after the initial run, the $J-V$ characteristics resemble that of an exponential set of traps distributed continuously in energy
(where $J \propto V^m $), as shown in Fig. 1. The slope of the trap-limited region increases till $m$=6, after which the $J-V$ characteristics
does not change for subsequent measurements. These additional traps are created by the injected carriers at high fields. We verified this by
sweeping the voltage of a new device (in Sample A) just below the trap-filled voltage and repeated the measurements several times; the $J-V$
characteristics does \emph{not} change at all as shown by the open symbol in the inset of Fig. 1 as long as the voltage is below $V_{TFL}$.
Again, after running the device once until trap-free SCLC, subsequent runs show the characteristics of a distribution of traps with $m$$>$2. It
is due to this generation of additional traps that the transition region between $V_{TFL}$ and trap-free SCLC deviates from being perfectly
vertical even in the first bias sweep. Moreover, the decrease in current at higher fields (trap-free SCLC region) suggests both emptying of
traps\cite{pope} and additional generation of trap states.

The inset of Fig. 2 shows the effect of detrapping by photoexcitation. The 457 nm line of an Ar$^+$ laser was used as the excitation source.  In
the figure, Run 5, 6 and 7 indicate $J-V$ characteristics before photoexcitation. The sample was illuminated for $\sim$ 15 min; remeasuring the
current for a forward bias sweep shows the Ohmic and the SCLC region quite clearly (as seen in the bottom curve of the inset). Although the
$J-V$ characteristics after photoexcitation reflect emptying of traps, it does not revert back to the initial shallow trap behavior. Since the
trap states are more localized comapred to the transport states, transitions from a localized trap state to the continuum are typically
forbidden in organic semicodnuctors because of the selection rules.\cite{schmechel} The incident light generates excitons (both singlet and
triplet), where the triplet excitons are also effective as detrapping agents.\cite{pope} Typically free carriers are generated by
autoionization, which then recombine with the trapped charges resulting in a detrapping process.\cite{schmechel} After photoexcitation, the
Ohmic region extends for higher bias values, suggestive of the introduction of free charge carriers by the photoexcitation process. Without
photoexcitation the lifetime of trapped charges is many hours. The exact nature of these long lifetimes is not known; it is being currently
investigated.

\subsubsection{Density of states calculation}\label{sec: DOS}
When traps are not discrete but have a distribution of energies, the slope of the $J-V$ curve is no longer 2 and changes continuously. The
density of states (DOS) distribution ($g(E_F$)) can be obtained by the differential method proposed by Nespurek and Swarakowski.\cite{nespurek}
In this scheme, the DOS at the quasi-Fermi level is given by
\begin{equation} \label{3}
g(E_{Fn}) = \left[ \frac{\chi \epsilon \epsilon_0}{ed^{2}kT} \right]
\left( \frac{V}{m(V)-1}\right),
\end{equation}
where $m(V)$ =$d(lnJ)/d(lnV)$.\cite{shubhra} $\chi$ is a correction factor for the non-uniformity of the internal field; here we take its value
as 1. The position of the quasi-Fermi level is $E_{Fn}=kTln(eN_{v}V\mu/Jd)$. This method is sensitive to the DOS at the quasi-Fermi level.

\begin{figure}
\unitlength1cm
\begin{picture}(4.0,7)
\put(-3,-1.){ \epsfig{file=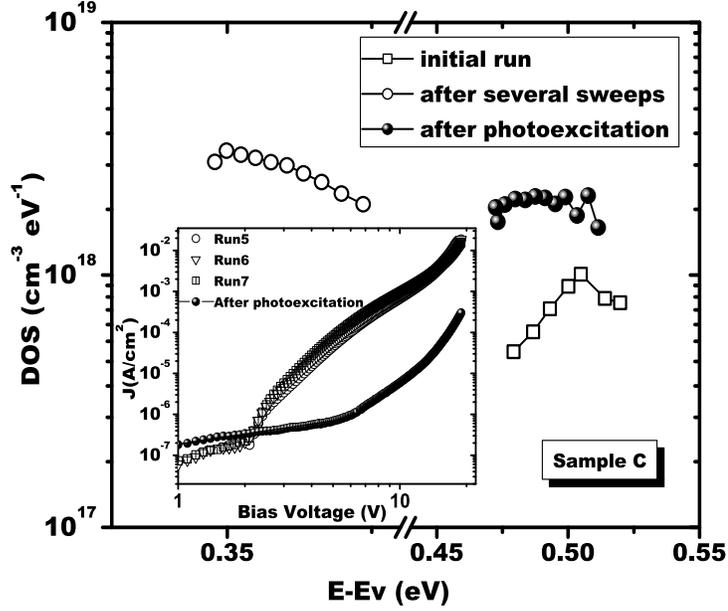, angle=0, width=11.cm,
totalheight=9.7cm}}
\end{picture}
\caption {DOS distribution as a function of the quasi-Fermi level
for the initial, trap-filled, and after photoexcitation bias sweeps
in Sample C. The inset shows successive $J-V$ curves after running
the device many times. The bottom curve in the inset was measured
after photoexcitation.} \label{figure2}
\end{figure}

Figure 2 shows the calculated DOS as a function of the quasi-Fermi level for three different situations in Sample C, using the method in Ref.
[\onlinecite{nespurek}]: (i) the very first bias sweep, which resembles the first run shown in Fig. 1; (ii) after several bias sweeps when the
traps are already filled (run 5, Fig. 2 inset); and (iii) after photoexcitation when the traps are emptied. Initially the DOS is
$\sim$7$\times$10$^{17}$ cm$^{-3}$eV$^{-1}$ centered at $\sim$0.5 eV. Since the DOS calculation provides the distribution of traps with respect
to the quasi Fermi energy level, traps created at high fields with several bias sweeps show their distribution as $\sim$2$\times$10$^{18}$
cm$^{-3}$eV$^{-1}$ at an energy of 0.35-0.4 eV. After photoexcitation the DOS remains at 2$\times$10$^{18}$ cm$^{-3}$eV$^{-1}$ with a shift of
the quasi-Fermi level to $\sim$0.5 eV, similar to the first run, suggesting partial emptying of traps. Since the DOS values remain higher
compared to the very first run even after the traps are emptied, it is a clear indication that additional traps are created at higher voltages.
Generation of trap states has also been observed by Lang \emph{et al.} in pentacene single crystals with bias-stress \cite{lang}; however, in
their work the DOS values are not calculated.

The origin of the initial set of traps could be structural or chemical defects, which typically give discrete trap states. As-is PF2/6 film
shows some signature of the hexagonal ordering at room temperature.\cite{tanto} The defects may occur at the grain boundaries in the
polycrystalline layers, resulting in shallow trap SCLC. The exponential set of traps that are created may arise from hydrogen- and
oxygen-induced defects.\cite{lang} PF2/6 is known for its fluorenone defects at the bridging carbon atom;\cite{scherf} application of high
voltages may create more of these defects.

\subsubsection{Temperature-dependent I-V}\label{sec: thermalcycled film}
Temperature-dependent $I-V$ measurements yield alternate methods to
estimate $N_t$ and $E_t$ by determining $\Theta$ as a function of temperature [see Eq. (2)]. Unfortunately, since the shallow-trap behavior is
seen only in the first bias sweep here, and driving the device above $V_{TFL}$ results in exponential trap levels, discrete trap SCLC model
cannot be used to extract $\Theta$ as a function of temperature. However, temperature-dependent $I-V$ measurements allow us to estimate the
density of traps induced at higher voltages. The temperature-dependent $I-V$ was measured after running the device to higher voltages for a few
times, ensuring a steady $I-V$ characteristics.

\begin{figure}
\unitlength1cm
\begin{picture}(4.0,7)
\put(-3,-1.){ \epsfig{file=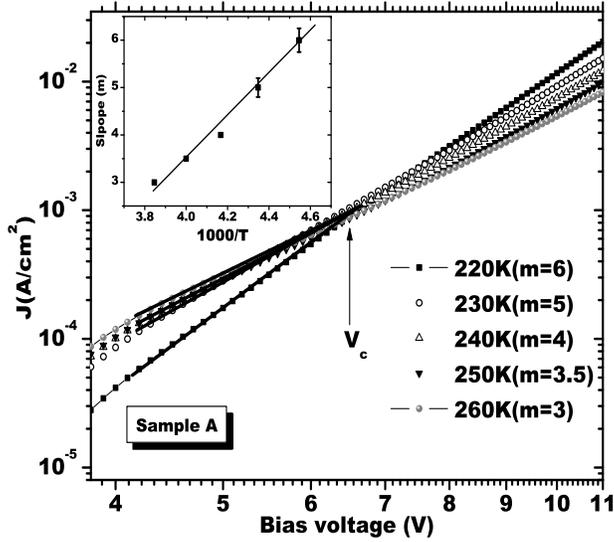, angle=0, width=10.cm, totalheight=8.7cm}}
\end{picture}
\caption{ $J$ vs. $V$ as function of temperature for Sample A. The cross-over voltage is at 6.5 V. The inset shows $m$ as a function of inverse
temperature.} \label{figure3 }
\end{figure}

Figure 3 shows $J$ vs.$V$ for Sample A measured at different
temperatures. Only a section of $J-V$ characteristics is shown to
highlight the voltage at which these curves cross each other. At the
cross-over voltage ($V_c$) the current is temperature independent,
denoting that the activation energy is zero. This voltage is related
to $N_t$ by $V_c =q N_{t} d^2/2\epsilon\epsilon_0$. \cite{rafiq}
Using $V_c$ = 6.5 V, we obtain $N_t$ as $2\times10^{17}$ cm$^{-3}$.
The inset of Fig. 3 shows that $m$ (obtained by fitting only a
narrow region of the \emph{J-V} curve) increases as a function of
inverse temperature. A similar temperature-dependent behavior is
observed for the other samples as well. The linear dependence of the
slope ($m$) with inverse temperature further establishes the
validity of the SCLC model for our PF2/6-based diodes.

\subsection{Thermal-cycled PF2/6 film}\label{sec: thermalcycled film}

\begin{figure}
\unitlength1cm
\begin{picture}(4.0,7)
\put(-3,-1.){ \epsfig{file=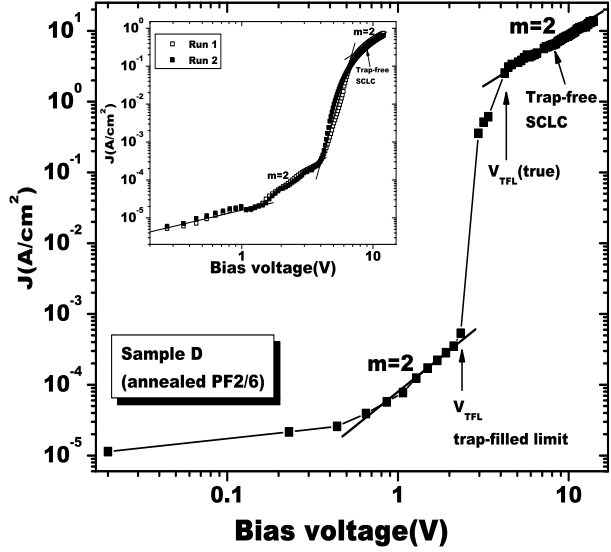, angle=0, width=10.cm, totalheight=8.7cm}}
\end{picture}
\caption{ $J$ vs. $V$ characteristics for Sample D. The inset shows two subsequent runs from another device of the same sample.} \label{figure3}
\end{figure}

Figure 4 shows the $J-V$ characteristics of Sample D, where the PF2/6 film is in a semicrystalline phase after annealing. The exact
crystallinity of these films has not been determined but our preliminary atomic force microscopy studies from annealed PF2/6 films show average
footprint area of $\sim$ 10$^4$ nm$^2$. Devices fabricated from this phase also show discrete traps under SCLC with all four regions. The
current density in the trap-limited and trap-free SCLC regions agrees very well with a $V^2$ dependence. The inset shows successive \emph{J-V}
runs from another device of the same sample; the slope of the trap-limited region no longer changes clearly indicating that no additional traps
are being created at higher voltages, when the PF2/6 film is in the semicrystalline form. These measurements were repeated for several devices
of Sample D and for all of them the \emph{J-V} characteristics are similar for successive runs with no additional creation of trap states.

\begin{table}
\caption{Hole mobility, trap density, and energy of the traps for
as-is and annealed PF2/6 devices.} \label{table2}
\begin{ruledtabular}
\begin{tabular}{cccccc}
Sample   & $\mu$(cm$^2$/Vs) &
$N_t$ (cm$^{-3}$) & $E_t$(eV)\\
\hline As-is    &  4$\times10^{-5}$-2.6$\times10^{-7}$ & 1.1-1.8$\times 10^{17}$  & 0.35-0.52\\
Annealed    &  2.3$\times10^{-4}$-2.1$\times 10^{-5}$  & 0.7-1.5$\times 10^{17}$ & 0.38-0.45 \\
\end{tabular}
\end{ruledtabular}
\end{table}
Table II compares the mobilities, density and energy of the trap states in as-is  and annealed (Sample D) PF2/6 devices. These results are from
more than 10 devices of each sample. The trap densities remain the same indicating that the structural or chemical defects in semicrystalline
PF2/6 is similar to the as-is polymer. The hole mobility is enhanced by more than an order of magnitude in the annealed sample. Such an
enhancement upon annealing has been reported in PF8-based devices.\cite{kreouzis} We point out that the as-is samples represented in this table
include Samples B and C (without the top PEDOT-PSS layer), which show slightly lower values of $E_t$. Although these samples behave as one
carrier hole-only devices there may be some electron injection resulting in slight variations.

A striking feature of Sample D is that additional traps are not
created upon subsequent bias sweeps unlike as-is PF2/6. Most likely
this reflects that the additional trap states arise from structural
disorder due to injected carriers at voltages higher than $V_{TFL}$.
In the hexagonal phase, the polymer chains adopt a specific
orientation preventing such disorders.

Discrete set of traps as seen in the \emph{J-V} characteristics for both as-is and annealed PF2/6 indicate that current injection is  primarily
governed by the ordered regions in the film and does not depend upon the overall bulk film properties. This is corroborated by our
capacitance-voltage measurements of PF2/6-based MIS diodes, where the equivalent parallel conductance demonstrate the presence of shallow trap
states with a single time constant.\cite{yun}

\section{Conclusions}\label{sec:conclusion}
In conclusion, the single trap level SCLC model for charge injection believed to be applicable only for crystalline materials of high chemical
and structural purity, is an excellent model for a class of blue-emitting polymers. PF2/6 with its high level of molecular disorder is an
exemplary system where the SCLC model within discrete level shallow traps can be applied to model current injection. These results shed new
light on the mechanism of charge injection and transport in polymers. Although  PF2/6 has a high degree of interchain disorder, the as-is
polymer shows regions of structural ordering that resembles the hexagonal phase \cite{tanto}; these ordered segments are most probably
responsible for charge injection and transport. The SCLC model with discrete shallow trap levels may be applicable to other amorphous polymers
which have regions of ordering. A cautionary remark is that the true SCLC behavior is observed only in the first bias sweep in the as-is
polymer, an effect that may be easily overlooked, since the injected carriers at voltages higher than $V_{TFL}$ generates a new set of
exponential traps.


\begin{acknowledgments}
We gratefully acknowledge the support of this work through the National Science Foundation under grant Nos. ECS-0523656 and DMR-0413601.

\end{acknowledgments}


\end{document}